# Design of Eutectic High Entropy Alloys

Ali Shafiei

Metallurgy Group, Niroo Research Institute (NRI), Tehran 14665-517, Iran
E-mail: alshafiei@nri.ac.ir, Tel: +98 (21) 88074788

**Abstract**

Eutectic high entropy alloys (EHEAs) are emerging as an exciting new class of structural alloys as they have shown promising mechanical properties. However, how to design these alloys has been a challenge. In this work, a thermodynamic approach based on the phase rule is developed for designing EHEAs. Furthermore, the ideas of compositional phase diagrams and mixing eutectic alloys are introduced, and it is shown how these methods could be used for designing new EHEAs. The approach is applied for alloy systems Al-Co-Cr-Fe-Ni and Co-Cr-Fe-Ni-Ti and several EHEAs are predicted for these alloy systems. The predicted results are verified with thermodynamic simulations and experimental data. The results show that the introduced approach can be considered as a feasible and easy-to-use method for designing EHEAs. Based on the developed approach, any binary or ternary eutectic alloy can be used for designing multicomponent eutectic alloys.

# 1. Introduction

High entropy alloys (HEAs) are a new class of metallic alloys which are founded on the idea that an alloy could have multiple (at least three) principal elements instead of one dominant element [1-2]. Based on this "fundamentally new idea" [3], new alloys, mostly selected from the central part of the multicomponent phase diagrams, are being designed and investigated; the aim is to identify alloys with enhanced properties in comparison with traditional alloys. Initially, the focus was on equiatomic alloys such as CoCrFeMnNi, CoCrFe, TiTaVMo, …, but currently, non-equiatomic alloys are also being investigated. The exploration has led to the discovery of some alloys with new physical phenomena and promising properties [3-5]. For example, one may name eutectic high entropy alloys (EHEAs) which are firstly reported by Lu et al. [6]. Due to their fine in-situ lamellar composite microstructures, EHEAs have shown very favorable combinations of strength and ductility [6–15] which have encouraged materials scientists to focus on EHEAs as a promising new class of structural alloys.

A challenge for further development of EHEAs is how to design these alloys. That is because the phase diagrams are not available for the majority of quaternary and quinary alloy systems. In fact, most of the EHEAs were found by trial and error methods. Several strategies are reported for designing EHEAs [9, 14, 16-22]. Although these methods where successful in designing EHEAs, they have their own limitations [19]. A simple approach was recently introduced by the author for designing EHEAs in Al-Co-Cr-Fe-Ni system [19]. It was assumed that EHEAs originate from binary and ternary eutectic systems [19]. As a result, the composition of binary and ternary eutectic alloys were used for finding the composition of EHEAs [19]. Furthermore, the concept of eutectic lines was introduced and it was proposed that new eutectic or near-eutectic compositions can be obtained by mixing the alloys which are located on the same eutectic line. In the present work, a thermodynamic approach is used to further explain the introduced model. Furthermore, more experimental data are provided to verify the model. The approach is then applied for alloy systems Al-Co-Cr-Fe-Ni and Co-Cr-Fe-Ni-Ti and new eutectic alloys are designed by using the introduced approach. The results show that the developed approach is a feasible and easy-to-use approach for designing EHEAs. Based on the developed approach, any binary or ternary eutectic alloy can be employed for designing multicomponent eutectic alloys.

## 2. Methodology

The ingots of designed AlCoCrFeNi and CoCrFeNiTi alloys were prepared via arc melting under a Ti-gettered high purity argon atmosphere. High purity constituent elements (Ni(99.9% wt.%), Co(99.99% wt.%), Al(99.999% wt.%), Cr (99.9% wt.%), and Ti (99.9% wt.%)) were used for making the alloys. The ingots were remelted four times to achieve compositional homogeneity. The homogenized ingot were then suction casted into 4 cm long and 8 mm diameter rods using a water-cooled copper mold. The samples were further sectioned perpendicular to thier length for microstructural investigations which were performed by optical microscope. Because the main objective of the present work is designing eutectic high entropy alloys, the goal was just observing eutectic microstructures. More detailed characterization techniques such as XRD and SEM were not performed for the characterization of samples. All of the simulations and thermodynamic calculations in the present work are performed by JMatPro® software version 7.0.0 developed by Sente Software Ltd. [23]. An alloy is considered as eutectic if simulation results predict that the solidification occurs in a narrow temperature range ($\Delta T_{max}$=10 °C), and if during the solidification simultaneous formation of two solid phases occurs [19, 38]. It should be noted that because a temperature range and not a fixed temperature is considered for the solidification of alloys, it can be interpreted that the alloys are in fact quasi-eutectic alloys.

## 3- Model development

In binary alloy systems, a binary eutectic reaction (e. g. L → γ + B2) can be considered as a three-phase equilibrium (liquid and two solid phases) with zero degree of freedom according to the phase rule (F=C−P+1, F=2-3+1=0). Therefore, a three-phase equilibrium appear as a point in a binary phase diagram (Figure 1a). In ternary alloy systems, a binary eutectic reaction becomes a three-phase equilibrium with 1 degree of freedom (F=3-3+1=1). Therefore, the traces of the phase compositions appear as lines in ternary phase diagrams (Figure 1b). In quinary systems, a binary eutectic reaction has three degree of freedom (F=5−3+1=3). Therefore, three-phase equilibria will form a hyper space within the quinary phase diagrams. This is schematically shown in Figure 1c for alloys with three-phase equilibrium L → γ + B2 in the quinary phase diagram Al-Co-Cr-Fe-Ni. Every alloy which is selected within this hyper space will go through a binary eutectic reaction during solidification. Therefore, if this hyper space can be defined, then designing eutectic alloys will be straight-forward, and for designing EHEAs, the objective should be defining this hyper space.

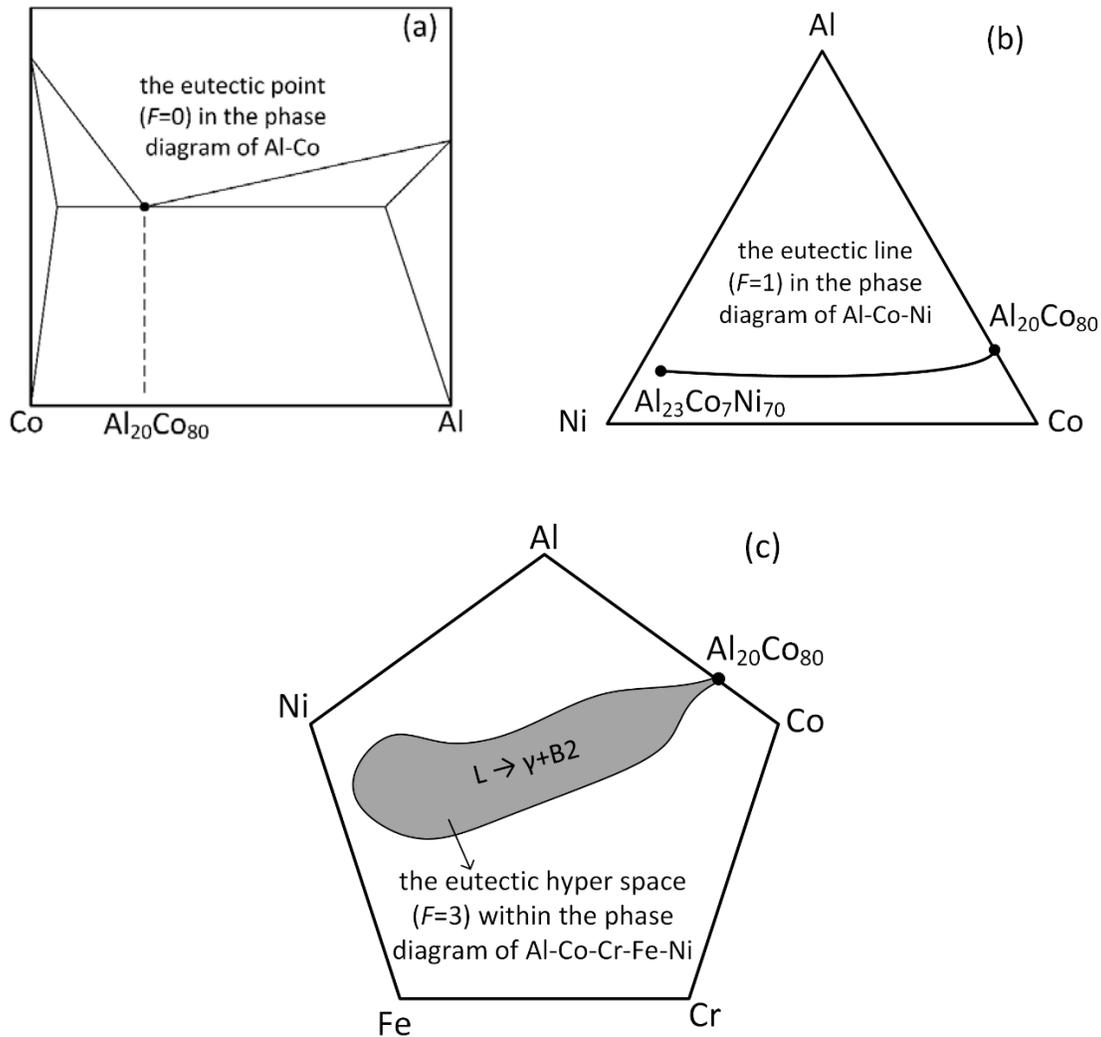

Figure 1. (a) The eutectic point in the phase diagram of Al-Co; (b) the eutectic line in the phase diagram of Al-Co-Ni; (c) the schematic of eutectic hyper space for alloys with three-phase equilibrium L → γ + B2 within the phase diagram of Al-Co-Cr-Fe-Ni system

As explained above, the traces of the liquid composition of three-phase equilibria form a hyper space within quinary phase diagrams. Such hyper spaces cannot be easily represented. Furthermore there are not enough experimental data to find these hyper spaces. One simple approach to show a hyper space could be finding the two dimensional projections of that hyper space. This is schematically shown in Figure 2 where a eutectic hyper space is projected on two projections. These two dimensional projections in fact define the limits of a eutectic hyper space within a phase diagram. Although, these projections cannot exactly define a eutectic hyper space, but they could be considered as good approximations for it and, therefore, they could be used for predicting the composition of EHEAs. If an alloy can be selected in a way to be located in these two dimensional projections, then with a great probability that alloy will be located within the eutectic hyper space and that alloy will be eutectic. This is the approach presented here for designing EHEAs. So the question now is how to find these projections. It is shown below that the compositions of binary and ternary eutectic alloys (which can be easily found from binary and ternary phase diagrams) can be used for finding the two dimensional projections of a eutectic hyper space. That is basically because binary and ternary eutectic alloys are also a part of that eutectic hyper space. This point is further explained below for the alloy system Al-Co-Cr-Fe-Ni.

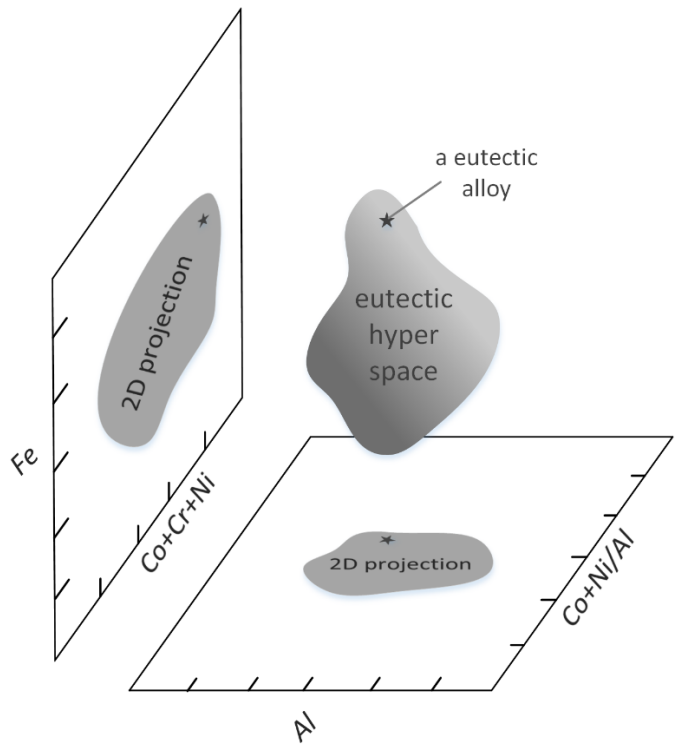

Figure 2. Schematic representing the projections of a eutectic hyper space on two different compositional diagrams

## 4. Al-Co-Cr-Fe-Ni system

The alloy system Al-Co-Cr-Fe-Ni is considered here first. This system is one of the extensively studied alloy systems in the field of HEAs. One of the eutectic reactions in this alloy system is L→ γ + B2. The chemical compositions of some of the verified binary and ternary eutectic alloys with this eutectic reaction are shown in Table 1. According to the ternary phase diagrams [26-30], alloys with this eutectic reaction are located along a line in ternary systems [26-30]; therefore ranges of compositions are listed for these ternary eutectic alloys in Table 1. Some of the experimentally verified EHEAs in this system are also listed in Table 2.

Table 1. The chemical compositions of some of the experimentally verified binary and ternary eutectic alloys in Al-Co-Cr-Fe-Ni system

| Alloy | Chemical composition (at.%) | | | | | Eutectic reaction |
|---|---|---|---|---|---|---|
| | Al | Co | Cr | Fe | Ni | |
| AlCo [34] | 20 | 80 | - | - | - | L→ γ + B2 |
| AlCoFe [31-32] | 15-20 | 63-80 | - | 0-22 | - | L→ γ + B2 |
| AlFeNi [26] | 17-22 | - | - | 10-50 | 33-68 | L→ γ + B2 |
| AlCoNi [29-30] | 20-23 | 7-80 | - | - | 0-70 | L→ γ + B2 |
| AlCoCr [33] | 19-20 | 56-80 | 0-25 | - | - | L→ γ + B2 |
| AlCrNi [27-28] | 17.5-23 | 0 | 4-31 | 0 | 51.5-73 | L→ γ + B2 |

Table 2. The chemical compositions of some of the experimentally verified EHEAs in Al-Co-Cr-Fe-Ni system

| Alloy | Chemical composition (at.%) | | | | | Eutectic reaction |
|---|---|---|---|---|---|---|
| | Al | Co | Cr | Fe | Ni | |
| AlCoFeNi [13] | 19 | 20 | - | 20 | 41 | L→ γ + B2 |
| AlCoCrNi [12] | 19 | 15 | 15 | - | 51 | L→ γ + B2 |
| AlCoCrNi [14] | 17.4 | 21.7 | 21.7 | 0 | 39.2 | L→ γ + B2 |
| AlCoCrNi [14] | 16 | 38.6 | 22.7 | 0 | 22.7 | L→ γ + B2 |
| AlCrFeNi$_2$ [35] | 16 | - | 20 | 20 | 44 | L→ γ + B2 |
| AlCoCrFeNi [36] | 18 | 30 | 10 | 10 | 32 | L→ γ + B2 |
| AlCoCrFeNi$_{2.1}$ [6] | 16.4 | 16.4 | 16.4 | 16.4 | 34.4 | L→ γ + B2 |
| AlCo$_2$CrFeNi$_2$ [18] | 17 | 28.6 | 14.3 | 14.3 | 25.8 | L→ γ + B2 |
| AlCoCrFe$_2$Ni$_2$ [18] | 17 | 14.3 | 14.3 | 28.6 | 25.8 | L→ γ + B2 |
| AlCoCrFeNi$_3$ [18] | 17 | 14.3 | 14.3 | 14.3 | 40.1 | L→ γ + B2 |
| AlCoCrFeNi [37] | 18 | 30 | 10 | 10 | 32 | L→ γ + B2 |
| AlCoFeNi [37] | 18 | 30 | - | 20 | 32 | L→ γ + B2 |
| AlCoFeNi [37] | 18 | 27.34 | - | 27.34 | 27.34 | L→ γ + B2 |
| AlCoCrFeNi [37]* | 18 | 30 | 10 | 10 | 30 | L→ γ + B2 |
| AlCoCrFeNi [37]* | 18 | 24 | 10 | 10 | 36 | L→ γ + B2 |
| AlCoCrFeNi [37]* | 18 | 20 | 10 | 10 | 40 | L→ γ + B2 |
| AlCoCrFeNi [38] | 16 | 41 | 15 | 10 | 18 | L→ γ + B2 |
| AlCo$_{0.8}$CrFeNi$_{2.3}$ [24] | 16.4 | 13 | 16.4 | 16.4 | 37.8 | L→ γ + B2 |
| AlCo$_{1.2}$CrFeNi$_{1.9}$ [24] | 16.4 | 19.7 | 16.4 | 16.4 | 31.1 | L→ γ + B2 |
| AlCoCr$_{1.2}$Fe$_{0.8}$Ni$_{2.1}$ [24] | 16.4 | 16.4 | 19.7 | 13 | 34.5 | L→ γ + B2 |
| AlCoCr$_{0.8}$Fe$_{1.2}$Ni$_{2.1}$ [24] | 16.4 | 16.4 | 13 | 19.7 | 34.5 | L→ γ + B2 |
| AlCo$_{0.6}$CrFe$_{1.43}$Ni$_{2.1}$ [24] | 16.3 | 9.8 | 16.3 | 23.3 | 34.3 | L→ γ + B2 |
| AlCo$_{1.2}$CrFe$_{0.81}$Ni$_{2.1}$ [24] | 16.4 | 19.7 | 16.4 | 13.2 | 34.3 | L→ γ + B2 |
| AlCo$_{1.2}$Cr$_{0.81}$FeNi$_{2.1}$ [24] | 16.4 | 19.7 | 13 | 16.4 | 34.5 | L→ γ + B2 |
| AlCoCrFeNi [24] | 18 | 30 | 10 | 10 | 32 | L→ γ + B2 |

* These alloys contained 2 at.% of W for improving the mechanical properties

Figure 3a shows the Al concentration versus (Ni+Co) concentration, and Figure 3b shows the Al concentration versus (Ni+Co+Fe) concentration of all eutectic alloys in Tables 1 and 2. Ternary eutectics are shown by lines in Figure 3. It can be seen that there are regions limited by binary and ternary eutectics in which all EHEAs are located. So it can be hypothesized that eutectic regions in Figure 3 are in fact two dimensional projections of the eutectic hyper space space within the phase diagram of Al-Co-Cr-Fe-Ni. Furthermore, it can be postulated that all EHEAs in Al-Co-Cr-Fe-Ni system should be located within these eutectic regions. Eutectic regions in Figure 3 provide references for evaluating the Al concentration of EHEAs in Al-Co-Cr-Fe-Ni.

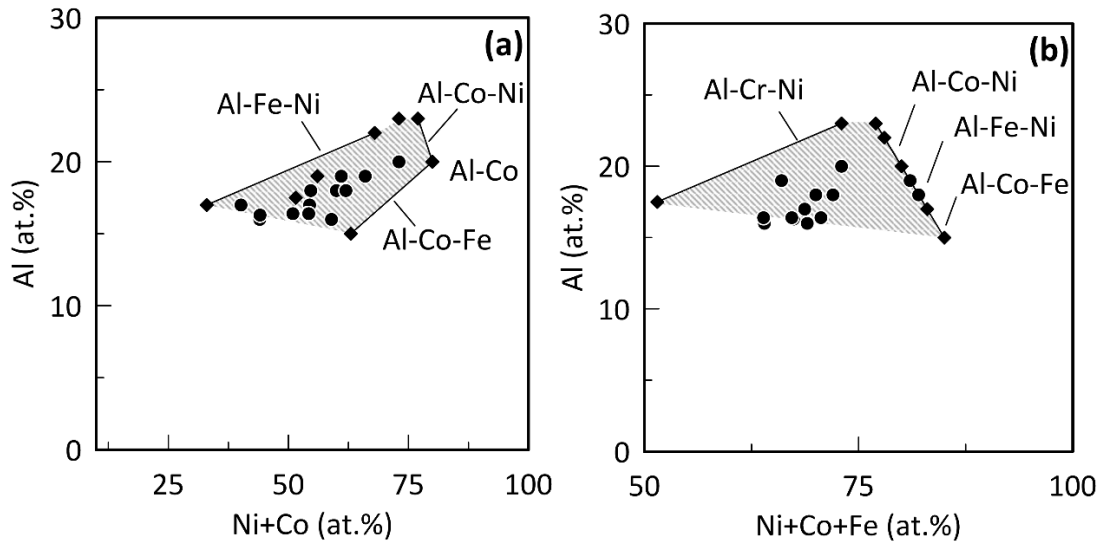

Figure 3. The Al concentration versus (a) (Ni+Co) and (b) (Ni+Co+Fe) concentration of binary and ternary (◆) and quaternary and quinary (●) eutectic alloys in Tables 1 and 2

Similar to the compositional diagrams in Figure 3, compositional diagrams in Figures 4 are plotted by using the concentration of eutectic alloys in Tables 1 and 2. According to the diagrams in Figure 4, again it can be seen that there are narrow regions limited by binary and ternary eutectic compositions in which all EHEAs are located. Furthermore, it can be anticipated that the Cr, Fe and Ni concentration of all (γ +B2) EHEAs in Al-Co-Cr-Fe-Ni system should be within the eutectic regions which are shown in Figure 4.

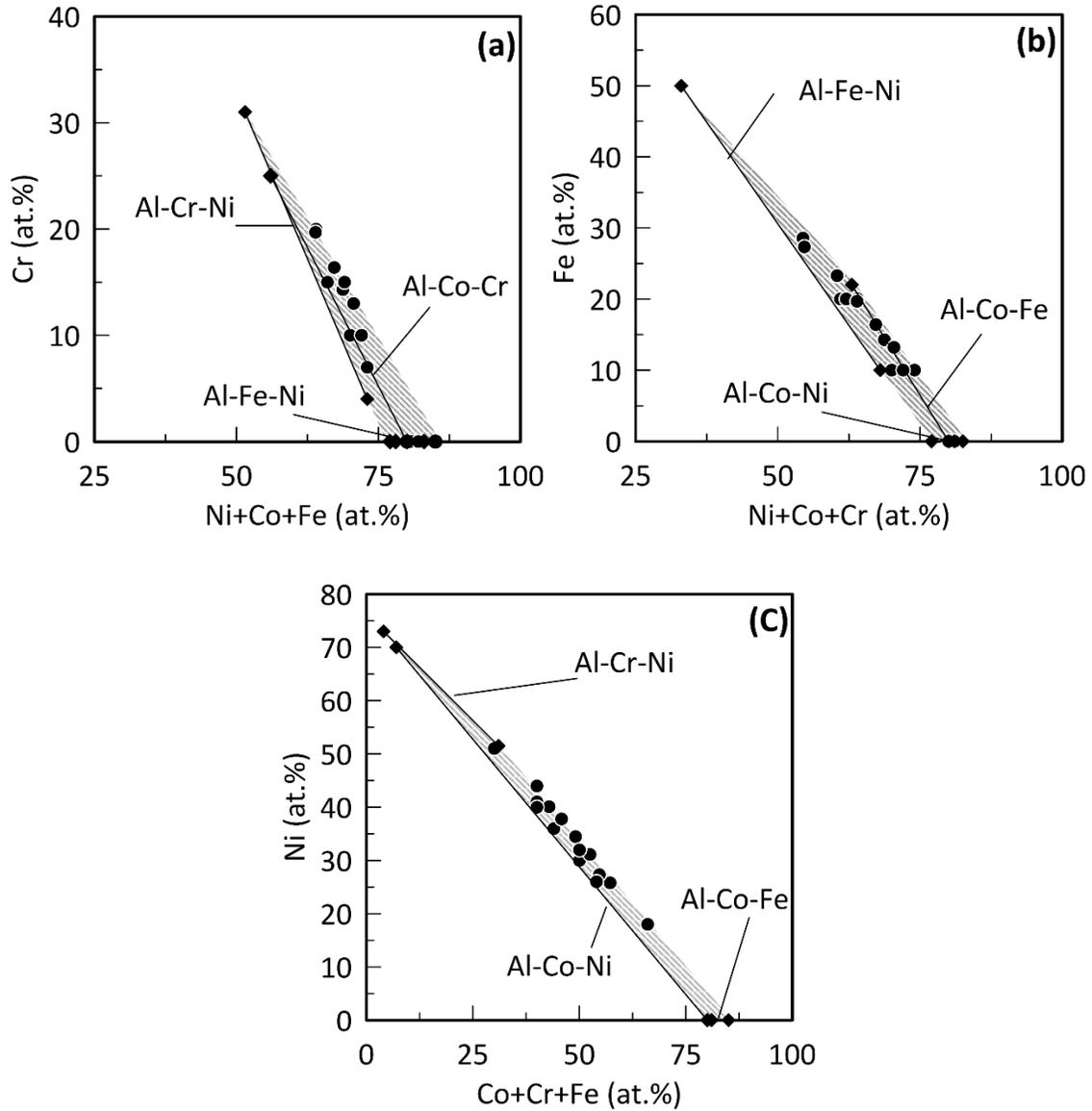

Figure 4. (a) The Cr concentration versus (Ni+Co+Fe) concentration, (b) the Fe concentration versus (Ni+Co+Cr) concentration and (c) the Ni concentration versus (Co+Cr+Fe) concentration of binary and ternary (◆) and quaternary and quinary (●) eutectic alloys in Tables 1 and 2

The eutectic regions in Figures 3 and 4 can be considered as two dimensional projections of the eutectic hyper space within the phase diagram of Al-Co-Cr-Fe-Ni. If an alloy can be designed in a manner to be located within all of the eutectic regions in Figures 3 and 4, the chance of that alloy to be located within the eutectic hyper space is very high and that alloys could be eutectic. Therefore, to design new EHEAs in Al-Co-Cr-Fe-Ni system, an alloy should be designed in a manner to be located in all of the eutectic regions shown in Figures 3 and 4.

For example, three alloys $Al_{18}Co_{24}Cr_7Fe_{17}Ni_{34}$, $Al_{18}Co_{43}Cr_{15}Ni_{24}$, and $Al_{18}Co_{34}Cr_{11}Fe_8Ni_{29}$ were designed. These alloys where designed in a way to be located within all of the eutectic regions in Figures 3 and 4. The thermodynamic simulation results for these alloys are shown in Figure 5. The simulation results clearly show that these alloys are eutectic indicating that the approach and the eutectic regions introduced in this work could be used for designing EHEAs. It should be emphasized that eutectic regions in Figures 3 and 4 are just approximately showing the eutectic hyper space within the phase diagram of Al-Co-Cr-Fe-Ni, and each alloy which is located inside of these eutectic regions is not necessarily eutectic. The alloys which are located in eutectic regions just have a good chance to be eutectic. If more compositional diagrams similar to Figures 3 and 4 could be designed and used for projecting the eutectic hyper space, then more accurate results will be obtained. In other words, if more compositional diagrams can be designed, then the eutectic hyper space can be projected more accurately and, as a result, more accurate eutectic compositions can be obtained.

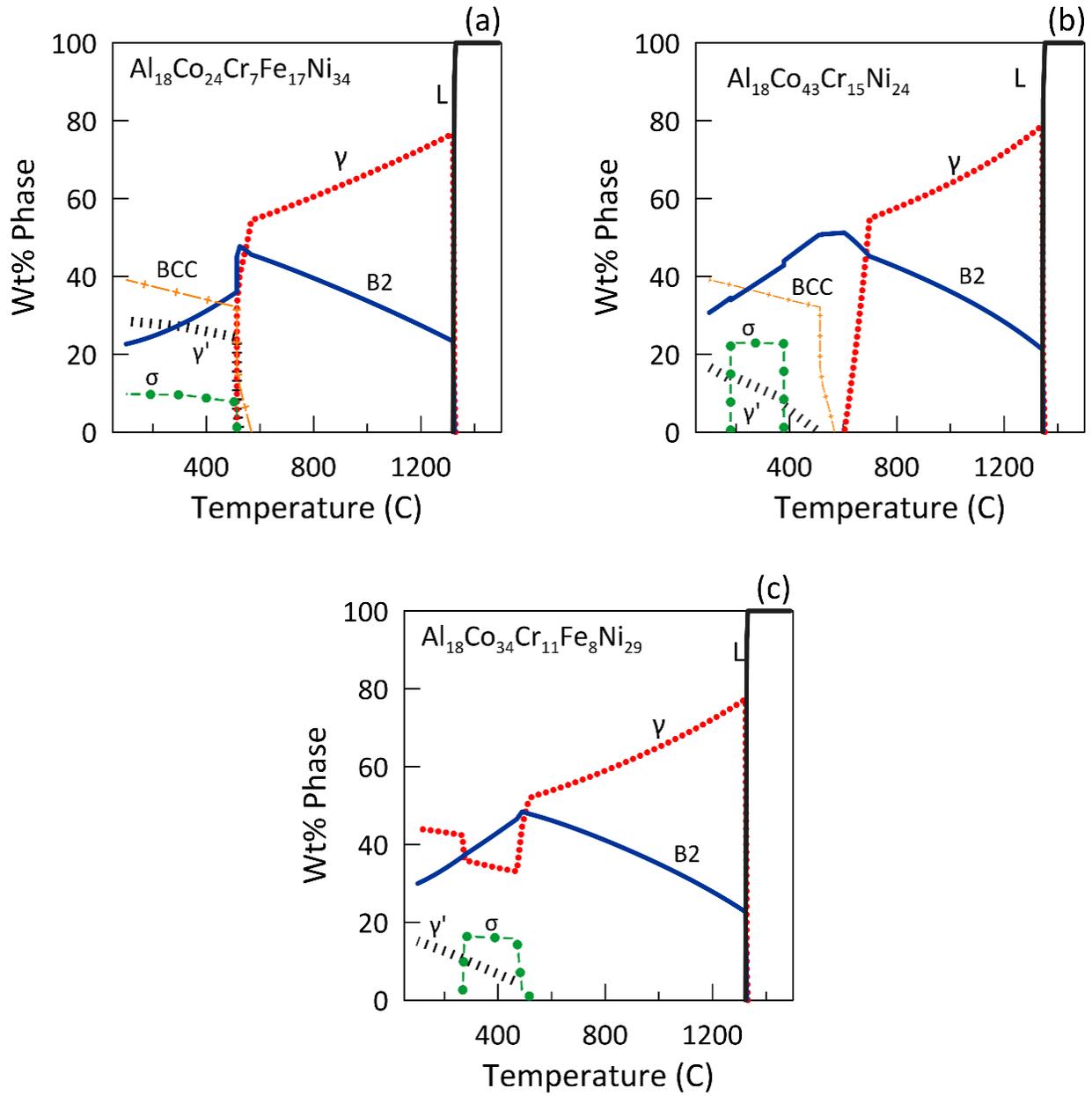

Figure 5. Thermodynamic simulation results for alloys (a) $Al_{18}Co_{24}Cr_7Fe_{17}Ni_{34}$, (b) $Al_{18}Co_{43}Cr_{15}Ni_{24}$ and (c) $Al_{18}Co_{34}Cr_{11}Fe_8Ni_{29}$

## 5- Mixing eutectic alloys for obtaining new eutectic alloys

A good estimation of a eutectic composition can be obtained by mixing eutectic alloys. That is because if two eutectic alloys are mixed together, the obtained alloy will also be probably inside the eutectic hyper space. This is schematically shown in Figure 6. Specifically, binary and ternary eutectic systems can be mixed for obtaining other eutectic compositions. That is because binary and ternary eutectic alloys form the boundaries of eutectic regions in Figures 3 and 4. For example, for designing a eutectic high entropy alloy in Al-Co-Cr-Fe-Ni, one may mix eutectic alloys $Al_{20}Co_{80}$ [34] and $Al_{19}Co_{15}Cr_{15}Ni_{51}$ [12] in a 1:1 molar ratio. The obtained composition is $Al_{19.5}Co_{47.5}Cr_{7.5}Ni_{25.5}$ which can be checked against the compositional diagrams in Figures 3 and 4 to see if it is located inside all of the eutectic regions or not. The idea of mixing eutectic alloys for obtaining other eutectic alloys was first reported in [19] and is based on the idea that EHEAs are originated from binary and ternary eutectic alloys and the point that eutectic alloys can be connected via eutectic lines. The existence of eutectic lines between eutectic alloys is also schematically shown in Figure 6.

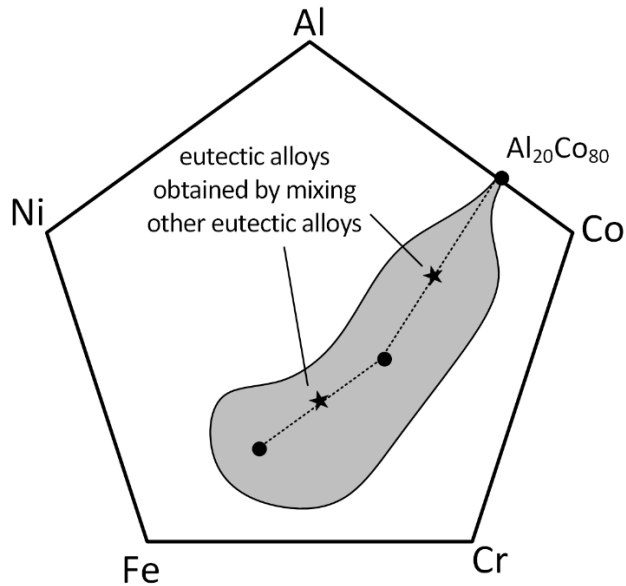

Figure 6. A schematic showing the existence of eutectic lines (dotted line) between eutectic alloys and obtaining new eutectic alloys (★) by mixing other eutectic alloys (●)

Alloy $Al_{19.5}Co_{47.5}Cr_{7.5}Ni_{25.5}$ designed above by mixing two eutectic alloys were made by casting. An optical image from the microstructure of alloy $Al_{19.5}Co_{47.5}Cr_{7.5}Ni_{25.5}$ is shown in Figure 7. It can be seen that the microstructure of alloy $Al_{19.5}Co_{47.5}Cr_{7.5}Ni_{25.5}$ consists from a fine intimate mixture of two phases indicating that alloy $Al_{19.5}Co_{47.5}Cr_{7.5}Ni_{25.5}$ is eutectic. So the approach was successful in predicting a new eutectic alloy composition.

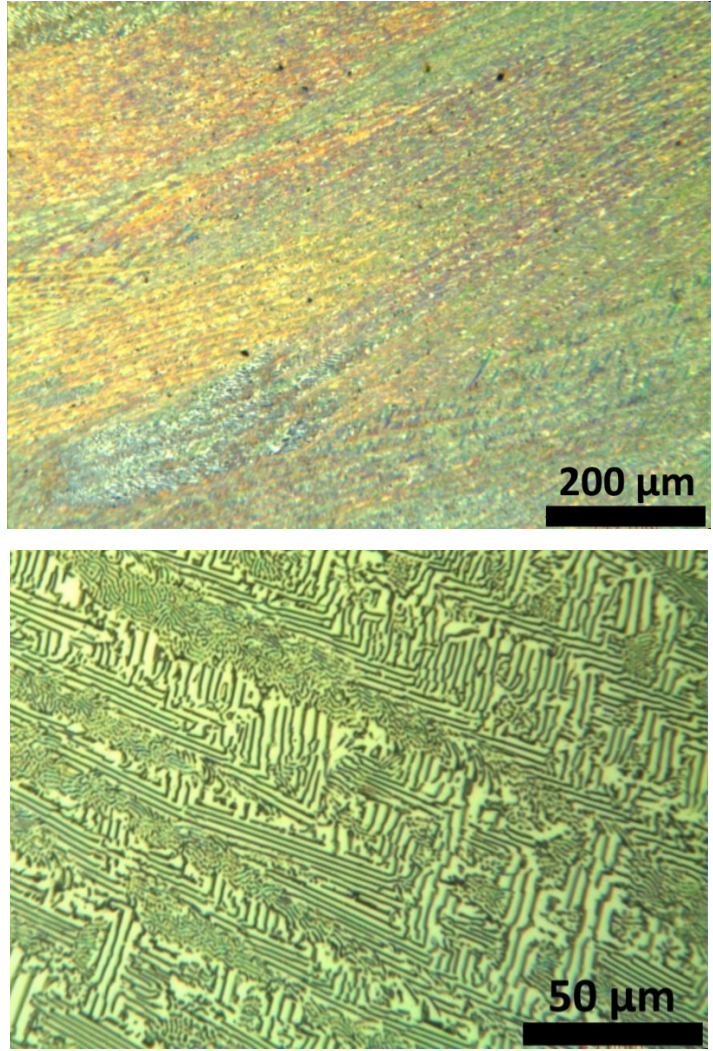

Figure 7. An optical image of the microstructure of alloy $Al_{19.5}Co_{47.5}Cr_{7.5}Ni_{25.5}$ designed by mixing eutectic alloys $Al_{20}Co_{80}$ [34] and $Al_{19}Co_{15}Cr_{15}Ni_{51}$ [12] in a 1:1 molar ratio

The approach presented above for finding the composition of EHEAs in Al-Co-Cr-Fe-Ni system, can be applied for any eutectic reaction in any alloy system. Therefore, the following steps can be proposed for designing EHEAs from a binary eutectic alloy:

1) Selecting a eutectic reaction and the associated binary eutectic alloy from a binary phase diagram (e. g. $Al_{20}Co_{80}$)

2) Selecting the elements which want to be added to the binary eutectic alloy (for example adding Ni, Fe, and Cr to $Al_{20}Co_{80}$ for developing AlCoCrFeNi EHEAs)

3) Finding the ternary eutectic alloys of that eutectic reaction from ternary phase diagrams

4) Plotting eutectic regions similar to the procedure used in Figures 3 and 4, and finally

5) Designing EHEAs by using the plotted eutectic regions (the first approximation can be obtained by mixing binary and ternary eutectic alloys)

It should be noted that if a more number of compositional diagrams can be used, then the eutectic hyper space will be projected more accurately and, as a result, more accurate eutectic compositions will be obtained. In the following section, the approach is applied for the alloy system Co-Cr-Fe-Ni-Ti.

**6- Co-Cr-Fe-Ni-Ti system**

To the best of the authors' knowledge, no EHEA is reported for this alloy system. Different binary eutectic reactions could be considered for this alloy system. One of the binary eutectic reactions is L→ γ + $Ni_3Ti$ which occurs in a binary alloy with the composition $Ni_{84}Ti_{16}$ [39]. This eutectic reaction can also be found in ternary alloy systems Cr-Ni-Ti [39], Co-Ni-Ti [40], and Fe-Ni-Ti [41] as they are shown in Figure 8. It can be assumed that there are alloys with eutectic reaction L→ γ + $Ni_3Ti$ in quinary alloy system Co-Cr-Fe-Ni-Ti. Alloys with this eutectic reaction will occupy a hyper space within the phase diagram of Co-Cr-Fe-Ni-Ti as it is schematically shown in Figure 9. Eutectic hyper spaces for two other eutectic reactions are also schematically shown in Figure 9. The eutectic reaction L→ γ + $Ni_3Ti$ is considered here.

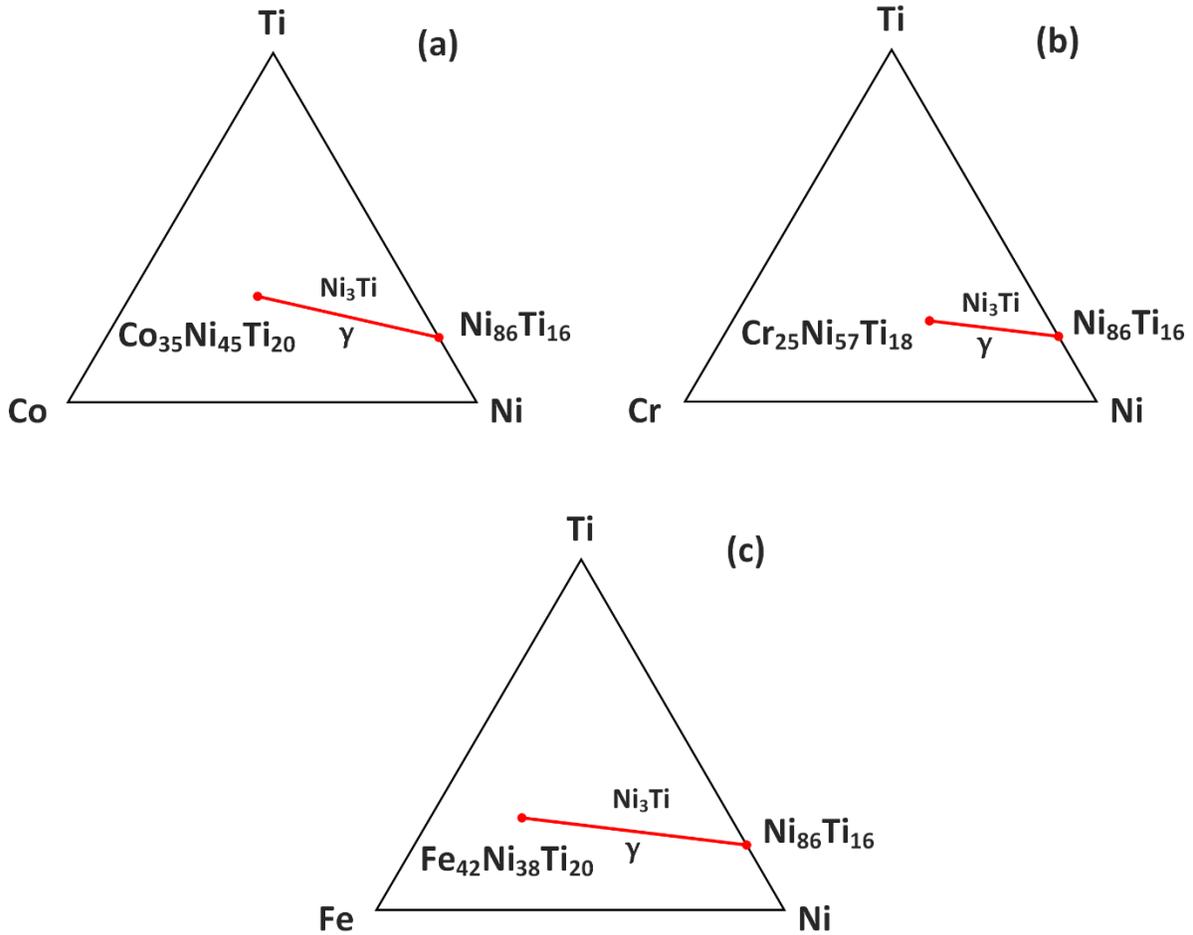

Figure 8. Schematics of ternary phase diagrams (a) Co-Ni-Ti [40], (b) Cr-Ni-Ti [39], and (c) Fe-Ni-Ti [41], and the eutectic lines for eutectic reaction L→ γ + $Ni_3Ti$

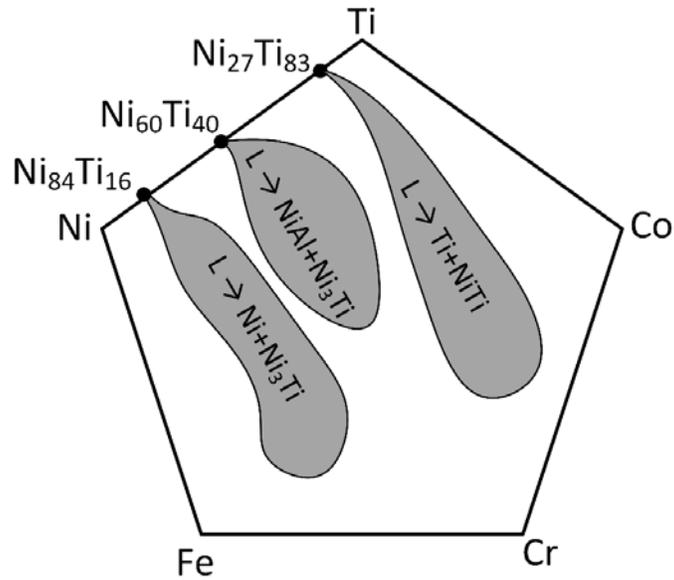

Figure 9. The schematic of eutectic hyper spaces within the phase diagram of Co-Cr-Fe-Ni-Ti system for three different eutectic reaction

By using the composition of binary and ternary eutectic alloys in Figure 8, the compositional diagrams in Figure 10 are plotted. Ternary eutectics are shown by lines in Figure 10. Dashed regions which are shown in Figure 10 could be considered as two dimensional projections of the eutectic hyper space related to the reaction L→γ+Ni$_3$Ti. Therefore, EHEAs with eutectic reaction L→γ+Ni$_3$Ti are expected to be located within the dashed regions in Figure 10. To design eutectic alloys, an alloy should be designed in a manner to be located inside all of the dashed regions in Figure 10. Alloy Co$_{11}$Cr$_7$Fe$_{11}$Ni$_{53}$Ti$_{18}$ was designed by mixing three eutectic alloys Ni$_{48}$Fe$_{33}$Ti$_{19}$ [41], Cr$_{21}$Ni$_{62}$Ti$_{17}$ [39], and Co$_{33}$Ni$_{47}$Ti$_{20}$ [40] in an eqiumolar ratio. The designed alloy is shown by a red circle in Figure 10. As it can be seen this alloy is located inside all of the dashed regions meaning that this alloy could be eutectic.

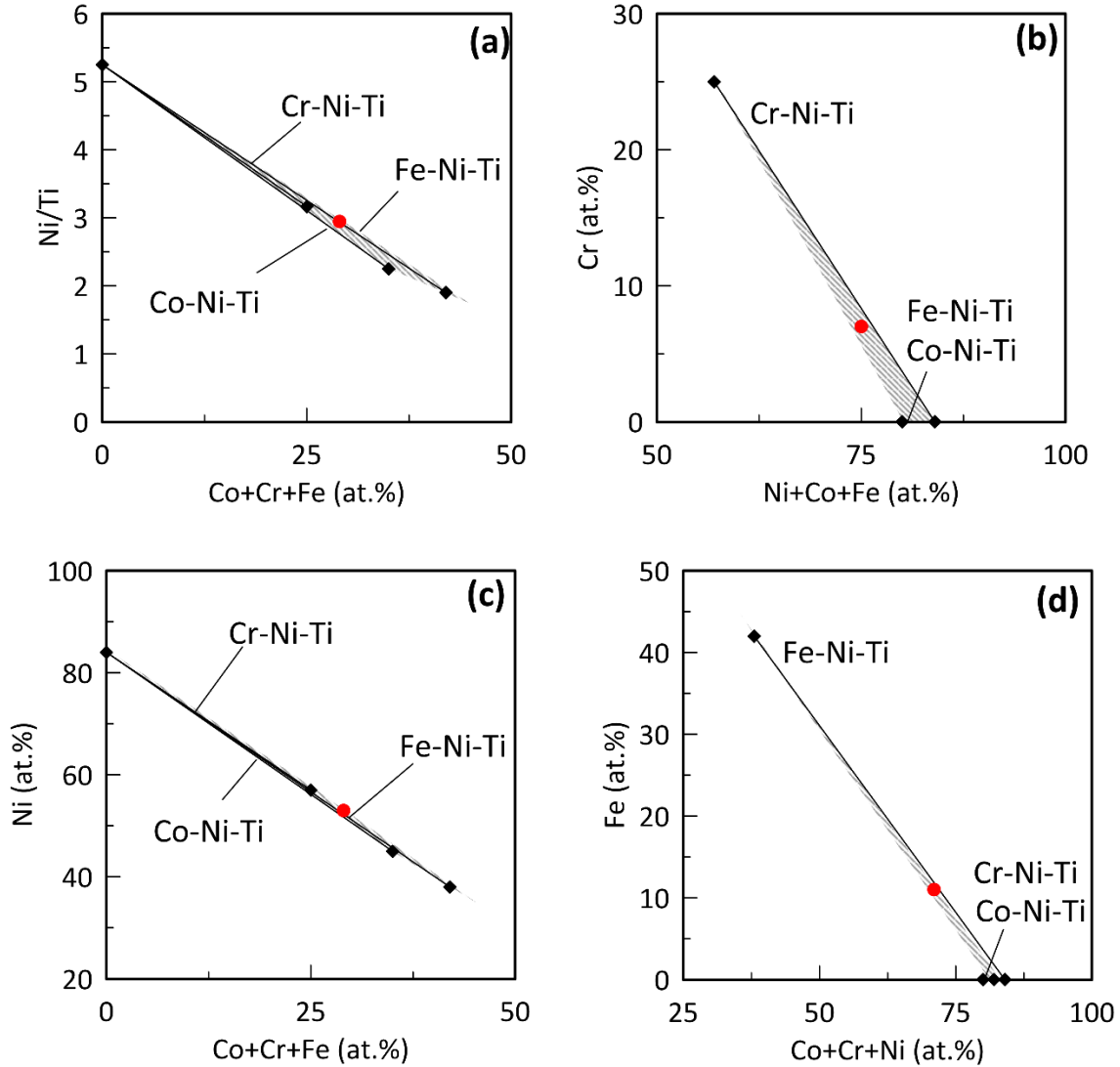

Figure 10. (a-d) Compositional diagrams for eutectic reaction L→ Ni+Ni$_3$Ti in quinary alloy system Co-Cr-Fe-Ni-Ti, and the designed eutectic alloy (●)

Alloy Co$_{11}$Cr$_7$Fe$_{11}$Ni$_{53}$Ti$_{18}$ was prepared via casting and the microstructure of this alloy is shown in Figure 11. It can be seen that the microstructure of the alloy consist of an intimate mixture of two phases indicating that the designed alloy is eutectic. So the approach presented here for designing eutectic alloys was successful.

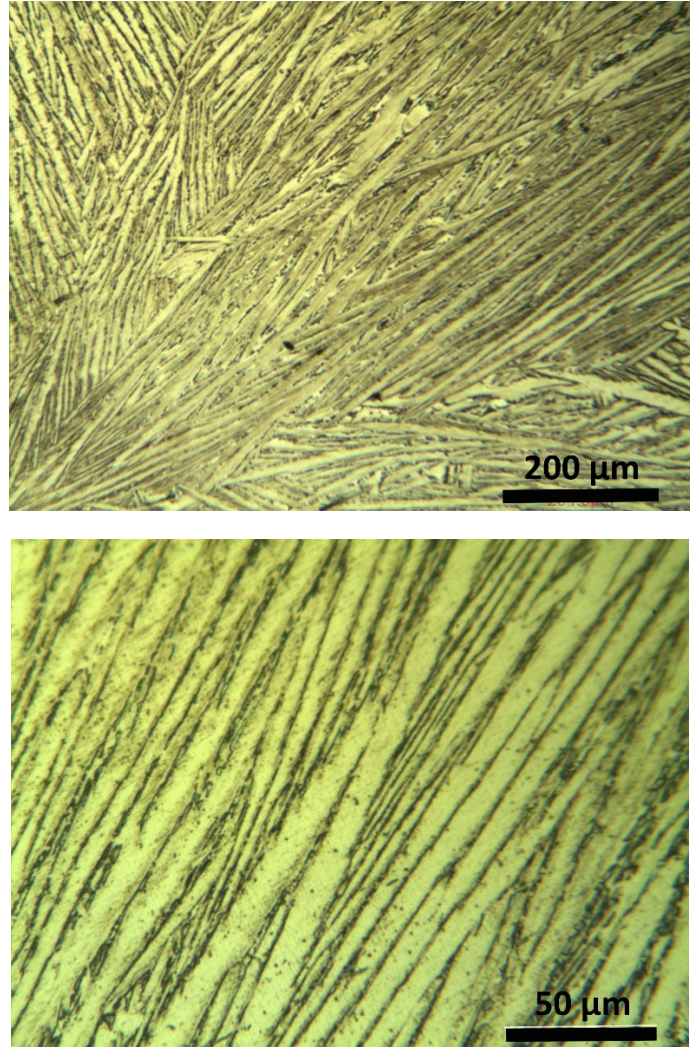

Figure 11. An optical image of the microstructure of alloy $Co_{11}Cr_7Fe_{11}Ni_{53}Ti_{18}$ designed by mixing eutectic alloys $Ni_{48}Fe_{33}Ti_{19}$ [41], $Cr_{21}Ni_{62}Ti_{17}$ [39], and $Co_{33}Ni_{47}Ti_{20}$ [40] in an eqiumolar ratio

## 7- Discussions

The thermodynamic phase rule indicates that liquid compositions of a three-phase equilibrium such as L → γ + B2 will form a hyper space within a quinary phase diagram. Therefore, if these hyper spaces can be defined, then designing eutectic alloys will be straight-forward. Compositional diagrams are introduced which can be used for projecting these hyper spaces and designing EHEAs. It is shown how these diagrams can be obtained by using the composition of binary and ternary eutectic alloys. These compositional diagrams reveal the composition range within which EHEAs can exist. Although these compositional diagrams cannot exactly define a eutectic hyper space within a phase diagram, they can be considered as good estimates of that eutectic hyper space. It should be emphasized that eutectic regions which are shown in compositional diagrams are just approximately showing the eutectic hyper space within a phase diagrams. Each alloy which is located inside of these eutectic regions is not necessarily eutectic. The alloys which are located in eutectic regions just have a good chance to be eutectic. If a more number of compositional diagrams are used for projecting a eutectic hyper space, then a more accurate projection of that eutectic hyper space will be obtained and, as a result, more accurate eutectic compositions can be obtained.

By investigating the compositional diagrams in Figures 3 and 4, it can be seen that eutectic regions, in which EHEAs are located, are limited by the composition of binary and ternary eutectic alloys. This observation suggests that a relation should exist between the composition of binary and ternary eutectic alloys and the composition of EHEAs. Furthermore, it can be postulated that EHEAs originate from binary and ternary eutectic alloys [19]. In other words, it can be hypothesized that EHEAs are made by adding new elements to binary and ternary eutectic alloys. For example, it can be assumed that (γ + B2) EHEAs in Al-Co-Cr-Fe-Ni system are made by adding elements Ni, Cr, and Fe to the binary eutectic alloy $Al_{20}Co_{80}$.

The idea of mixing eutectic alloys and the compositional diagrams introduced in this work can be used together for designing EHEAs in any alloy system. In general, they provide simple approaches for designing EHEAs. Two alloy systems are considered here but the approach can be extended to other alloy systems. The author applied this approach for alloy systems Co-Cr-Fe-Ni and Co-Cr-Fe-Ni-Ta and the results can be found in [42]. An interesting case could be eutectic refractory high entropy alloys which are reported very recently [43 and 44]. Another application of the developed approach could be designing EHEAs with low melting temperatures which could have possible applications in brazing and soldering. An alloy system which could be suggested in this regard is Al-Ag-Bi-Sn-Zn. Further research and experiments are needed in this regard.

## 8- Conclusions

The thermodynamic phase rule indicates that a eutectic hyper space should exist for a three-phase equilibrium reaction such as L $\rightarrow$ γ + B2 in quinary phase diagrams. Compositional diagrams can be used for defining these eutectic hyper spaces within quinary phase diagrams. These compositional diagrams clearly show that a connection exists between the composition of binary and ternary eutectic alloys and the composition of EHEAs implying the point that EHEAs originate from binary and ternary eutectic alloys. It is shown how these compositional diagrams can be used for designing new eutectic alloys. The idea of mixing eutectic alloys is introduced in this work, and it is shown how by mixing binary and ternary eutectic alloys new eutectic alloys can be designed. By using the ideas of compositional diagrams and mixing eutectic alloys, several eutectic high entropy alloys are designed for alloy systems Al-Co-Cr-Fe-Ni. Furthermore, for the first time a eutectic high entropy alloy ($Co_{11}Cr_7Fe_{11}Ni_{53}Ti_{18}$) is designed for Co-Cr-Fe-Ni-Ti system. The designed eutectic alloys are verified with thermodynamic simulations and experimental data. The results show that the idea of mixing eutectic alloys and the compositional diagrams can be successfully used together for designing EHEAs. Based on the developed approach, any binary eutectic alloy can be used for designing multicomponent eutectic alloys.

# 9- References